# Simulating cellular communications in vehicular networks: making SimuLTE interoperable with Veins

Giovanni Nardini, Antonio Virdis, Giovanni Stea
Dipartimento di Ingegneria dell'Informazione
University of Pisa, Italy
g.nardini@ing.unipi.it, a.virdis@iet.unipi.it, giovanni.stea@unipi.it

*Abstract*— The evolution of cellular technologies toward 5G progressively enables efficient and ubiquitous communications in an increasing number of fields. Among these, vehicular networks are being considered as one of the most promising and challenging applications, requiring support for communications in high-speed mobility and delay-constrained information exchange in proximity. In this context, simulation frameworks under the OMNeT++ umbrella are already available: SimuLTE and Veins for cellular and vehicular systems, respectively. In this paper, we describe the modifications that make SimuLTE interoperable with Veins and INET, which leverage the OMNeT++ paradigm, and allow us to achieve our goal without any modification to either of the latter two. We discuss the limitations of the previous solution, namely VeinsLTE, which integrates all three in a single framework, thus preventing independent evolution and upgrades of each building block.

## I. Introduction

Over the last two decades, two independent research areas have witnessed huge progress: cellular communications, on one side, and vehicular networks, on the other. Parallel to increasing the end-user bandwidth and providing ubiquitous coverage, research on cellular networks has also defined and analyzed a growing number of use cases, incorporating their requirements in the latest standards. This is especially evident with the onset of 5G research, a key issue of which is providing guaranteed performance to the largest possible number of users, regardless of their position. This requires the network to be deployed in a denser manner, to support high-speed mobility in a reliable manner, to perform fast handover between neighboring cells, etc.. Parallel to this, and quite likely due to the above improvements, research on vehicular technologies has started to consider cellular networks as a promising communication technology. On one hand, cellular communications allow existing vehicular network services to be improved, and new ones to be enabled. On the other hand, recent research projects envision vehicular communications as a promising use case for cellular systems, under the name of "connected cars" [1], [2].

The interactions between the above two fields can be seen from two main points of view: first, it can enable the so-called vehicle-to-everything (V2X) communications, including communications among vehicles (V2V) or between cars and the infrastructure (V2I), i.e. scenarios where vehicles are either or both end-points of the communication. Second, vehicular mobility can be seen as a specific case of user mobility, thus introducing new challenges to the underlying communication network, such as bulk handovers, unexpected network congestions during traffic jams, etc.

Network simulation has been widely used in both fields to test network architectures, algorithms, communication protocols, and applications on a large scale. In the frame of OMNeT++, there are two frameworks that tackle vehicular networks and cellular systems, called Veins [3] and SimuLTE [4][5], respectively. Veins provides vehicular mobility to OMNeT++, using *SUMO* [6] as the underlying vehicular-traffic simulator. SimuLTE, instead, is a system-level-simulator of LTE networks, based on the INET framework. Although the two frameworks are developed for the same simulation system, they cannot be used together "as-is". A first integration attempt has been made with VeinsLTE [7], which integrates a customized version of both in a single package. However, this solution defines a third standalone framework, rather than interconnecting the two, taking snapshots of two independent ongoing developments. This makes it difficult, if possible at all, to upgrade it when the composing framework gets upgraded. For instance, the recent additions of device-to-device communications in SimuLTE [8],[9] are not automatically subsumed in VeinsLTE. Dually, the addition of the support to platooning [10] in Veins does not automatically enable it in VeinsLTE.

In this paper, we describe the process of making SimuLTE and Veins *interoperable*, i.e. using both in the same simulation scenarios with the specific aim of keeping them separate and independent. We highlight the requirements coming from Veins and we list the main modifications to SimuLTE required to satisfy these requirements, in particular detailing how we manage dynamic creation and destruction of LTE-capable nodes. The rest of this paper is organized as follows: Section II provides the background on SimuLTE. In Section III we describe the requirements coming from Veins and the related work on the topic. Section IV, then, describes the integration process and Section V provides an example of simulation configuration. Finally, Section VI concludes the paper.

## II. Background on SimuLTE

In this section, we provide a background on the main architectural elements of SimuLTE. The latter is a framework for system-level simulation of LTE and LTE-Advanced cellular networks. It is based on OMNeT++ and exploits the INET framework to implement all the higher layers of the IP stack as well as the main IP nodes of the communication network, such as IP routers and application servers. Moreover, INET provides the concept of Network Interface Card (NIC) modules, which can be included within other modules to implement models of various communication protocols between network devices. As an example, devices can be empowered with Wi-Fi or Point-to-Point Protocol (PPP) communication capabilities by integrating the respective NIC modules. SimuLTE exploits the concept of modularity coming from OMNeT++ to realize its main building block: the LTE NIC card. The latter is designed as an extension of a wireless NIC module from INET, and allows one to add LTE capabilities to a node included in the simulation.

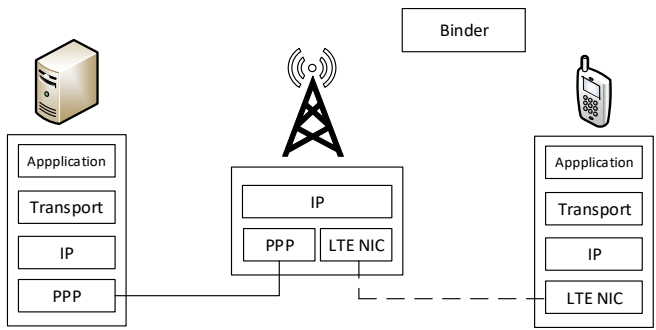

Figure 1 - High-level representation of an LTE scenario using SimuLTE

SimuLTE provides models for the two main elements of the LTE communication architecture, namely the user equipment (UE) and the evolved NodeB (eNB). Both nodes contain the LTE NIC, as shown in Figure 1, together with modules implementing upper layer protocols, taken from INET. The eNB has also an interface to the Internet via PPP, which provides means to send/receive data traffic coming from application servers, implemented as INET *standard hosts.* Multiple eNBs can also be connected together using the X2 interface, enabling coordination algorithms such as CoMP [11]. Within the LTE NIC located at both the UE and the eNB, SimuLTE implements a complete LTE protocol stack by means of a submodule for each layer, namely Packet Data Convergence Protocol (PDCP), Radio Link Control (RLC), MAC and PHY. SimuLTE exploits the inheritance paradigm of OMNeT++ to define both structure and behavior of UEs and eNBs. According to this paradigm, each layer will have common operations implemented in a base class. The latter is then extended to implement functionalities specific for the UE and the eNB, respectively. Both communication between adjacent layers and data transmission between UEs and eNBs are implemented via message exchange. Accounting of transmission resources is managed separately from data transmission, i.e. transmission signals are not explicitly modeled. To do so, the *Binder* module keeps track of how many and which transmission resources, i.e. Resource Blocks (RBs), are used during each transmission slot of 1ms. Two distinct sets of RBs are used respectively by the eNBs for downlink transmissions, and by the UEs for uplink transmissions. The Binder thus acts as an oracle for the LTE network: all the LTE nodes in the simulated network register to it at the beginning of the simulation, and can access it via direct method calls to share or obtain common information. For instance, this allows us to model control-plane operations with negligible cost in term of complexity.

The ChannelModel class included within the PHY layer of each LTE NIC models the effects of the air channel on transmissions between nodes. Whenever a new message reaches a NIC, the corresponding ChannelModel evaluates the Signal-to-Interference-and-Noise Ratio (SINR) as perceived by the node. This computation accounts for the intercell interference, which affects LTE communications. To do so, the channel model on the receiving end obtains information from the Binder regarding the RB allocation for all the nodes in the network. The compute SINR is then used to decide whether or not the message can be successfully decoded, depending on its value and on the modulation and coding scheme (MCS) used for that transmission. The ChannelModel is also responsible for the computation of the UE's Channel Quality Indicator (CQI). The latter is used by the eNB to select the MCS used for transmission, and might be also used for scheduling operations at the eNB, e.g. to enforce maximum-throughput allocation schemes. SimuLTE provides a realistic implementation of the ChannelModel, which takes into account path loss and fading effects. Such implementation can be easily extended or replaced, thus allowing researchers to use whatever physical-layer model they see fit.

SimuLTE allows one to simulate both communication between UEs and eNBs, and among UEs. If two UEs are both served by the same eNB, they can communicate either using the latter as a relay or in a direct manner, called *device-to-device* (D2D) communication [12]. Moreover, the LTE NIC provided by SimuLTE can be used to empower any network device with LTE communication capabilities. This way it is possible to design nodes with multiple and heterogeneous network interfaces, e.g. LTE, Wi-Fi, Bluetooth, etc.

III. VEINS REQUIREMENTS AND RELATED WORK

In this section, we describe the requirements that SimuLTE must meet in order to integrate vehicular mobility provided by the Veins framework. Then, we give an insight on the limitation of the existing solution, namely *VeinsLTE* [7].

Support to vehicular mobility within Veins is provided by *SUMO* [6], a popular simulation tool that allows users to create detailed road traffic scenarios. The responsibility of interacting with SUMO through the *TraCI* interface [8] is given to a module called *TraCIScenarioManager*, which must be present in any network definition. In particular, the TraCIScenarioManager obtains the information about the movements of vehicles in the simulated road traffic scenario and updates their mobility information within OMNeT++. To do this, vehicles (defined as OMNeT++ compound modules) need to be equipped with a *TraCIMobility* submodule. For vehicular mobility to work with SimuLTE, UEs need to incorporate this type of mobility. However, in SimuLTE, mobility of UEs is handled by the INET framework and its latest versions do not provide TraCIMobility. This means that modifications are required to add this support. Vehicles can also enter and/or leave the simulation dynamically, i.e. they can appear/disappear at any time during the simulation. This is done by the `addModule` and `deleteManagedModule` functions implemented by the TraCIScenarioManager. The `addModule` function instantiates the new OMNeT++ module implementing the vehicle and calls its initialization procedure. The `deleteManagedModule`, instead, invokes the `finish` function of the module and removes it from the simulation. These operations are accomplished by using the OMNeT++ API. If the new module is an LTE-capable node, it needs to be registered to the Binder of SimuLTE, i.e. stored in the data structures of the Binder. Similarly, on deletion, it must be removed from the Binder's data structures. Moreover, when considering cellular communications, a newly created vehicle must be associated with its serving base station (i.e. the eNB in case of the LTE network). Typically, vehicular networks cover large geographical areas (e.g., a city), hence a *multicell* environment must be considered. Since the new vehicle can appear at any position of the simulation playground, a procedure that performs the attachment of the incoming vehicle to the best serving cell is necessary and this should occur during the initialization phase of the module. For example, the vehicle can be associated to the eNB from which it perceives the highest value of the Signal to Interference and Noise Ratio (SINR), although this is not the only possible association criterion. In any case, the best serving eNB can change during the simulation, due to mobility and/or varying conditions in the environment (e.g. interference). Thus, support for *handover* (HO) must be available too.

The first attempt to integrate SimuLTE and Veins has been made by *VeinsLTE* [7]. The latter provides a single package that puts together Veins, SimuLTE and INET. However, it does so by making modifications to both Veins and SimuLTE, such that they need to interact *directly* with each other. For example, registration to the Binder during the creation of new vehicles in the `addModule` function is performed via direct method calls. The same occurs when removing vehicles in the `deleteManagedModule` function. This is a problem if the data structures of the Binder are changed in future releases of SimuLTE, e.g. to add new functionalities. Moreover, the eNB to which the new vehicle has to be associated is defined statically by the `ini` file, with no possibility of selecting the best serving cell at runtime. As far as mobility is concerned, the INET version included in the VeinsLTE package (i.e. INET v2.3) provides the TraCIMobility modules, hence the only thing to do is to configure that mobility for LTE-capable nodes in the `ini` file. However, INET removed such support recently.

IV. MODIFICATIONS TO SIMULTE

In this section, we describe the main modifications to the code of SimuLTE to allow the interoperability with Veins.

Figure 2 shows the *Car* module implemented within SimuLTE, which is similar to the structure of a legacy UE. With respect to the latter, a *vehicularMobility* module has been added. This is defined as an interface that can be implemented by the TraCIMobility module defined by Veins. Note that the *INETMobility* module is still present, so as to maintain backward compatibility. Since the Car cannot use both mobility modules during the simulation, they are defined as conditional modules within the `ned` file. Anyway, both modules exploit the mechanism of *signals* provided by OMNeT++ to trigger mobility events, which are handled by modules that are *subscribed* to them. According to the activated mobility module, the Car subscribes itself to the signals generated by the corresponding module. Unlike with VeinsLTE, it is not necessary that the TraCIMobility module implements the INET mobility interface.

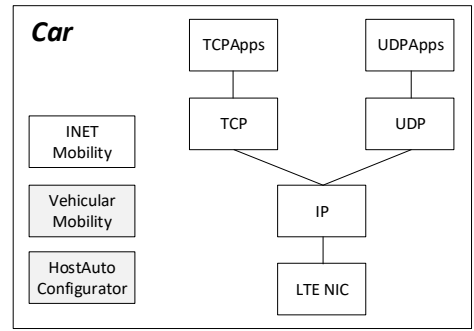

Figure 2 – Car module definition

As far as dynamic creation of vehicles is concerned, Veins uses the OMNeT++ API to create and initialize the new module, as mentioned in Section III. However, previous releases of SimuLTE were designed so that modules' initialization was only performed at the beginning of the simulation. For this reason, we reorganized the initialization functions within the entire LTE protocol stack of the Car module, so that SimuLTE takes care of all the initialization related tasks of the LTE NIC, thus allowing the registration to the Binder at any time. Similarly, when a vehicle leaves the simulation, Veins just invokes the latter's `finish()` function, which in turn calls the `finish()` function for all the submodules of the vehicle to be removed, if implemented. Thus, we defined the behavior of those functions for the relevant submodules of the LTE NIC so as to deregister the vehicle from the Binder and clear the buffers (at both UE and eNB side).

When a new vehicle is created, it needs to obtain an IP address to communicate. SimuLTE demands the assignment of IP addresses to the *IPv4NetworkConfigurator* module provided by INET. However, this module performs the assignment only at the beginning of the simulation, hence it cannot handle modules created dynamically. To solve the problem, we endowed the Car with a *HostAutoConfigurator* module. The latter is a deprecated module that INET's authors have left with the specific purpose of enabling the addition of network nodes at runtime. New vehicles must also be associated to one eNB according to a given criterion. To do so, we allow the vehicle to measure the power received from every simulated eNB and to perform the attachment to the one with the largest value of received power, something that was not possible using previous SimuLTE versions. Mobility across different cells is handled by the existing HO mechanism provided by SimuLTE, which is transparent to the fact that the UE can be a vehicle. We remark that no modifications to Veins have been made to enable the above functionalities, making them independent from the employed version of the Veins framework, as long as the requirements described in Section III hold.

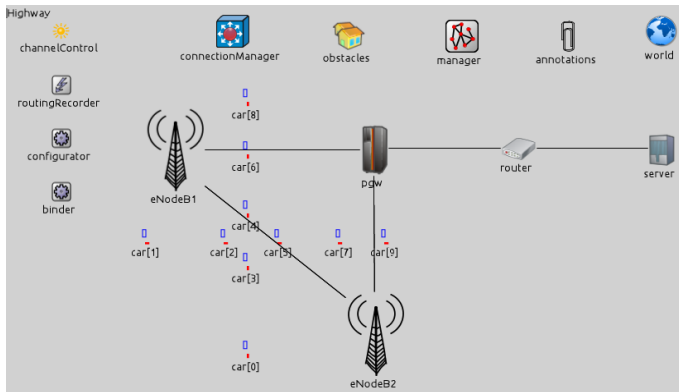

Figure 3 - Simulation scenario

## V. SIMULATING VEHICULAR NETWORKS

In this section, we show how to configure a simulation that integrates SimuLTE and Veins in order to evaluate cellular communications in a vehicular environment. We describe only the (few) relevant parameters in the `ini` configuration file needed to accomplish this task, whereas we refer the reader to the exemplary configurations provided by SimuLTE and Veins frameworks to tune their respective parameters. Similarly, the configuration of the specific road traffic scenario is performed by SUMO and this is outside the scope of this paper.

An example of network definition is shown in Figure 3. It contains ten vehicles created dynamically and two eNBs connected via the X2 interface, and to a remote server through a simplified Evolved Packet Core (EPC) network. The definition of the network must comprise the Binder and the TraCIScenarioManager modules (*binder* and *manager* in Figure 3, respectively). The latter needs to be informed about the type of the vehicles it needs to add to the simulation. This is accomplished by setting the following parameters.

```
*.manager.moduleType="lte.corenetwork.nodes.cars.Car"
*.manager.moduleName="car"
```

The `moduleType` parameter defines the path of the `ned` definition of the Car module we described in Section IV. The `moduleName` parameter, instead, states that we can configure parameters related to the new modules by referring to them using the name `car`. For example, we can impose that the first added vehicle (namely, `car[0]`) encounters an accident after 20 seconds from its departure time and that the accident lasts 30 seconds. As the snippet of code below shows, this is done by modifying the parameters of the `vehicularMobility` module of `car[0]`.

```
*.car[0].vehicularMobility.accidentCount = 1
*.car[0].vehicularMobility.accidentStart = 20s
*.car[0].vehicularMobility.accidentDuration = 30s
```

Initial association of vehicles to the best serving eNBs is disabled by default, but it can be activated as follows:

```
**.dynamicCellAssociation = true
```

Otherwise, manual association must be specified as follows:

```
*.car[*].masterId = 1
*.car[*].macCellId = 1
```

In this example, all vehicles are associated to eNodeB1 during their initialization, regardless of their position. In any case, HO can be activated by specifying `**.enableHandover = true`.

## VI. CONCLUSIONS AND FUTURE WORK

In this paper, we described the integration of SimuLTE and Veins, with the specific goal of preserving them as independent frameworks. We first detailed the requirements coming from Veins to allow custom nodes to be managed according to its mobility model. We then described the modification we implemented on SimuLTE to fit the above requirements, thus creating scenarios where LTE capable nodes are moved according to vehicular patterns. Finally, we provided an example of network configuration, highlighting the main configuration parameters.